\documentclass[graybox]{svmult}

\usepackage[bottom]{footmisc}

\usepackage{amssymb}
\usepackage{amsmath}
\usepackage{bm}

\usepackage{newtxtext}       %
\usepackage[varvw]{newtxmath}       

\usepackage{color}

\def\beq{\begin{equation}}
\def\eeq{\end{equation}}
\def\nn{\nonumber}

\def\om{\omega}
\def\a{\alpha}
\def\b{\beta}
\def\g{\gamma}
\def\th{\theta}

\def\s{\sigma}
\def\D{\Delta}

\def\p{\phi}

\def\r{\rho}
\def\l{\lambda}

\def\ah{\hat{a}}
\def\ad{\ah^\dagger}

\def\rd{{\textrm{d}}}
\def\paz{{\partial_z}}

\def\ra{\rightarrow}

\def\Cc{\mathbb{C}}

\def\Rr{\mathbb{R}}

\def\ph{\phantom}

\def\H{\mathcal{H}}

\def\B{{\cal B}}

\def\bz{\bar{z}}

\newcommand{\braket}[2]{\langle #1 \vert #2 \rangle}

\def\Ar{\hat{A}_k^{red}}

\begin{document}

  \title*{The $k$-photon quantum Rabi model}

  \author{Daniel Braak}
  \institute{Daniel Braak \at TP III and
   Center for Electronic Correlations and Magnetism,\\
Institute of Physics,
University of Augsburg, 86135 Augsburg, Germany
 \email{daniel.braak@uni-a.de} }
    
\maketitle

 \abstract{
 A generalization of the quantum Rabi model is obtained by replacing the linear (dipole) coupling between the two-level system and the radiation mode by a non-linear expression in the creation and annihilation operators, corresponding to multi-photon excitations. If each spin flip involves $k$ photons, it is called the ``$k$-photon'' quantum Rabi model. While the formally symmetric Hamilton operator is self-adjoint in the case $k=2$, it is demonstrated here that the Hamiltonian is not self-adjoint for $k\ge 3$. Therefore it does not generate a unique unitary time evolution and is unphysical. This result cannot be obtained by numerical calculations in finite-dimensional spaces which attempt to approximate an unbounded operator by a finite-rank operator.}




 \section{Introduction}
 \label{sec-intro}
  The quantum Rabi model with linear coupling is a well understood elementary model of quantum optics (see e.g. \cite{larson2021}) with Hamiltonian
  \beq
  H_{R}=\om\ad\ah + g\left(\ah + \ah^{\dagger}\right)\s_x +\D\s_z,
  \label{hamr}
  \eeq
coupling a bosonic mode (described by bosonic creation/annihilation operators $\ah,\ah^\dagger$) to a two-level system or ``pseudospin'', described by the Pauli matrices $\s_x,\s_z$. One sees from \eqref{hamr} that a spin flip, generated by $\s_x$, leads to a concomitant emission (effected by $\ah^\dagger$) or absorption ($\ah$) of a single light quantum. One would therefore assume that the following generalization of $H_R$, 
\beq
H_{kp}=\om\ad\ah + g\left(\ah^k + \ah^{\dagger k}\right)\s_x +\D\s_z,
\label{hamk}
\eeq
corresponds to a physical process in which each spin flip is accompanied by the emission or absorption of $k$ photons. Indeed, processes involving $k$-photon resonance have been observed and are of great interest for various application in quantum optics \cite{akeweje2021,menard2022}, but those processes are not described by \eqref{hamk}, but by the $k$-photon Jaynes-Cummings model,
\beq
H_{JCk}=\om\ad\ah + g\left(\ah^k\s^+ + \ah^{\dagger k}\s^-\right) +\D\s_z,
\label{hamjck}
\eeq
with $\s^\pm=(\s_x\pm i\s_y)/2$, the spin raising/lowering operators. The Hamiltonian $H_{JCk}$ is easily diagonalizable and has analytical properties completely different from $H_{kp}$. It can be directly derived for each experimental implementation if the coupling $g$ is small enough \cite{klimov2001,baghshahi2015}. In this sense $H_{JCk}$ is \emph{not} the ``rotating-wave'' approximation of $H_{kp}$, in contrast to the case $k=1$, where the Jaynes-Cummings model appears as the weak coupling (and close to resonance) approximation of $H_R$ which in turn is obtained directly from the dipole limit of the atomic light-matter interaction \cite{larson2021}. 

Nevertheless, the model $H_{kp}$ has been studied from a theoretical and mathematical viewpoint for several years, especially the case $k=2$ \cite{duan}, which features the ``spectral collapse'' phenomenon for sufficiently strong coupling, $g>g_c=\om/2$. For values $g<g_c$ the model can be physically implemented via trapped ions \cite{felicetti}. For the case $g>g_c$, it was conjectured in \cite{ng-99} that $H_{2p}$ is not self-adjoint: It is easily seen that for $g>\om/2$ and $\Delta=0$, the Hamiltonian \eqref{hamk} can be mapped to a harmonic oscillator with inverted potential. Therefore, it has no ground state and indeed no eigenvectors, the pure point spectrum is empty. However, the harmonic oscillator with inverted potential is known to be self-adjoint \cite{wienholtz,hellwig,kalf} and the continuous spectrum spans the whole real axis, similar to the position operator $\hat{x}$. The same applies to  $H_{2p}$ \cite{braak-23}. The case $k\ge 3$ has been studied in \cite{lo-98} and \cite{zhang2017}. Lo \emph{et al.}, as well as Zhang, come to the conclusion that the operator $H_{kp}$ is ``ill-defined''. While Zhang only states that the pure point spectrum is empty, which does not exclude self-adjointness, it is correctly claimed in \cite{lo-98} that $H_{kp}$ is not self-adjoint for $k\ge3$. To show this, Lo {\it et al.} use Nelson's theorem in the reverse, i.e. it is assumed that an operator $\hat{A}$ is not self-adjoint if some dense set of vectors $\{\p_n\}$ in the Hilbert space $\H$ yields no convergence of the expression
\beq
\sum_{j=0}\frac{t^j}{j!}\parallel\hat{A}^j\p_n\parallel
\label{nelson}
\eeq
for any $t>0$ and all $\p_n$. Nelson's theorem states that a symmetric operator $\hat{A}$ is essentially self-adjoint if $\mathcal{D}(\hat{A})$ contains at least one such  set of vectors, which are then called analytic vectors for $\hat{A}$ \cite{reed-2}. Therefore, in order to prove the negative ($\hat{A}$ is not essentially self-adjoint), one has to show that for \emph{all} dense sets $\{\p_n\}$ in  $\mathcal{D}(\hat{A})$ the expression \eqref{nelson} does not converge. It is not sufficient to prove it for just one set (in \cite{lo-98} this was shown for the eigenvectors of the harmonic oscillator). Indeed, the usual way to show lack of self-adjointness for $\hat{A}$ is constructing the domain of $\hat{A}^\dagger$, which is larger than $\mathcal{D}(\hat{A})$ if $\hat{A}$ is symmetric, a highly non-trivial task \cite{wienholtz, reed-1}.

In the present paper, it is proven with Bargmann space methods and asymptotic analysis that $H_{kp}$ has a continuous spectrum (in fact, all of $\Cc$), while the eigenvectors are normalizable\footnote{This seems to contradict Zhang's statement in \cite{zhang2017}. But Zhang's scalar product differs from \eqref{bargs}, so it cannot be compared to the present calculation.}. However, these elements of $L^2(\Rr)$ are not located  $\mathcal{D}(\hat{A})=\mathcal{D}(\ah^k)\cap\mathcal{D}(\ah^{\dagger k})$.  


$H_{kp}$ is formally symmetric because it contains the sum of operators $\ah^k$ and $\ah^{\dagger k}$ which are mutually adjoint in $L^2(\Rr)$. The total Hilbert space of $H_{kp}$ is $\Cc^2\otimes L^2(\Rr)$. In the following, we shall study the problem in the Bargmann space of analytic functions $\p(z)$, $\B$, which is isomorphic to $L^2(\Rr)$ \cite{bargmann,vourdas}. In this space, $\ah$ is given by the derivative $\paz$ and $\ah^\dagger$ by multiplication with $z$. The operators $z$ and $\paz$ are mutually adjoint in $\B$ with the scalar product  
\beq
\braket{\p}{\psi} =\frac{1}{\pi}\int \rd z\bz\ e^{-z\bz}\bar{\p}(\bz)\psi(z)
\label{bargs}
\eeq
A function $\p(z)$ is an element of $\B$ if $\p(z)$ is analytic in $\Cc$ and $\braket{\p}{\p}<\infty$.
After a trivial transformation of \eqref{hamk} which interchanges $\s_x$ and $\s_z$, we obtain the eigenvalue equation (setting $g=1$),
\begin{align}
  \left(\om z\paz +z^k +\paz^k\right)\p_1(z) +\D\p_2(z) &= E\p_1(z),\nn\\
  \left(\om z\paz -z^k -\paz^k\right)\p_2(z) +\D\p_1(z) &= E\p_2(z),
  \label{schrod}
\end{align}
for an element $\big(\p_1(z),\p_2(z)\big)^T$ in $\Cc^2\otimes\B$.

One could attempt now a shortcut by studying the decoupled system by setting $\D=0$, arguing that $\D$, as a bounded operator, cannot influence the self-adjointness properties of $\H_{kp}$. But we shall see in the following that in this way only part of the actually needed information (namely only one of the exponents of first kind, see below) is gained. Therefore, we shall first eliminate $\p_2(z)$
to obtain an equation for $\p_1(z)$,
\beq
\left[(\om z\paz -E)^2 -\left(z^k +\paz^k\right)^2 + \om k\left(z^{k-1}-\paz^{k-1}\right)
  -\D^2\right]\p_1=0.
\label{redeq}
\eeq
Normal ordering of the term $\left(z^k +\paz^k\right)^2$ yields
\beq
\left(z^k +\paz^k\right)^2=\paz^{2k} +z^{2k}+2z^k\paz^k+
\sum_{j=1}^k a_j^{(k)}z^{k-j}\paz^{k-j},
\label{normord}
\eeq
with known functions $a_j^{(k)}$. Equation \eqref{redeq} is an ordinary differential equation in the complex domain with no singular points in $\Cc$ except at $z=\infty$, where it has an unramified irregular singular point of $s$-rank three and class $2k$ \cite{slavyanov,braak-23}. Therefore all formal solution of \eqref{redeq} are analytic in $\Cc$ with an essential and isolated singularity at $z=\infty$. Thus, $\p_1(z)$ will be an element of $\B$ if it is normalizable with respect to the measure \eqref{bargs}. The formal solutions of \eqref{redeq} have 
the
asymptotic expansion for $|z| \ra\infty$,
\beq
\psi(z)=e^{\frac{\g}{2}z^2+\b z}z^\r
\sum_{n=0}^\infty c_nz^{-n}
\label{asym}
\eeq
for all $k\ge2$ \cite{braak-23}. In the following sections we shall use the method applied in \cite{braak-23} to the case $k=2$ to study $H_{kp}$ for all integer $k\ge 3$.

\section{ The cases $\bm{k=3,4}$   \label{sec-34}}

In this section, we treat the cases with $k=3,4$ separately, because the exponents behave differently for these low values of $k$ from the case $k\ge 5$, studied in section \ref{sec-k5}. Especially, they depend on $\om$ (albeit not on $E$ as for $k=2$ \cite{braak-23}).

\subsection{$\bm{k=3}$ \label{sec-k3}}
A direct calculation yields for the coefficients $a^{(3)}_j$,
$a_1^{(3)}=9$, $a_2^{(3)}=18$, $a_3^{(3)}=6$.
Then \eqref{redeq} reads for $k=3$,
\begin{align}
\left[
  -\paz^6 -2z^3\paz^3  +(\om^2-9)z^2\paz^2 +(\om^2-2E\om-18)z\paz \right.\nn\\
  \left.-z^6 +3\om z^2+E^2-\D^2-6\right]\p_1=0.
\label{redeq3}
\end{align}
The equation for the exponent of second kind with order two, $\g$, is obtained by plugging the expansion \eqref{asym} into \eqref{redeq3},
\beq
\g^6+2\g^3-1=(\g^3+1)^2=0.
\eeq
There are three different, each doubly degenerate values for $\g$, located on the unit circle (sixth roots of unity). This entails that the solutions of \eqref{redeq3} are conditionally normalizable, determined by the subdominant exponents
$\b$ and $\r$ which lift the remaining two-fold degeneracy of the asymptotic expansion. A direct calculation yields for $\b(\g)$,
\beq
\b=\pm\frac{\om}{3\g}.
\eeq
The two possible values of $\b$ for each $\g$ give altogether six linearly independent asymptotic forms for the formal solution of \eqref{redeq3}. Each such solution is only valid in a certain sector of the complex plane due to the Stokes phenomenon \cite{slavyanov}. Which expansion is asymptotically valid in a given Stokes sector cannot be computed from the values of $\p_1(z)$ in any bounded domain of the complex plane. However, we shall see that the condition of normalizability does not depend on the knowledge of the Stokes phenomenon.
The critical line for the exponent $\g$ with $\vert\g\vert=1$, where the integral in $\braket{\p_1}{\p_1}$ is not controlled by the factor $\exp(-z\bz)$ in the Bargmann measure \eqref{bargs} is given by the ray $z(r)=r\exp(-i\th/2)$ with $\th=\textrm{arg}(\g)$ \cite{braak-23}. The solution is normalizable if for all $\g$ the $\g$-critical lines are not in the Stokes sector where $\g$ is valid. On the other hand, if the critical line for some $\g$ lies in the Stokes sector of $\g$, we find  
\[
\p_1(z) \sim \exp\left(\frac{1}{2}r^2 \pm \frac{\om}{3}e^{-i3\th/2}r\right) z^\r \big(c_0+\mathcal{O}(r^{-1})\big).
\]
Because $\textrm{arg}(\g)\in\{\pi,\pm\pi/3\}$, the factor $\exp(\b z)$ has unit modulus along the critical line and does not determine the finiteness of $\vert\p_1\vert$. The convergence of the integral in $\braket{\p_1}{\p_1}$ is then given by the value of $\r(\g,\b)$. The explicit calculation yields $\r(\g,\b)=-2$ for all possible $\g,\b$. Because $\Re(\r)<-1/2$, it follows that $\vert\p_1\vert<\infty$. This is true whether a Stokes sector contains a critical line or not, thus knowledge of the Stokes multipliers is not needed, as proven in \cite{braak-23}. This means that the eigenvalue equation has normalizable solutions for all $E\in\Cc$ and all six combinations of $\g$ and $\b$. This entails that $H_{3p}$ is not self-adjoint, although it may have self-adjoint extensions because the defect indices are $\{6,6\}$ \cite{reed-2}. However, these extensions are not unique and it is not possible to define the extended domain of $H_{3p}$ by restricting the behavior of $\p_1(z)$ for $z\ra\infty$ in a physically meaningful way. For example, just demanding a finite photon content, $\langle\p_1\vert\ah^{\dagger}\ah\vert\p_1\rangle \stackrel{!}{<}\infty$, does not remove enough of the uncountably many, not mutually orthogonal eigenstates.

\subsection{$\bm{k=4}$ \label{sec-k4}}
In this case, we have $a_1^{(4)}=16$, $a_2^{(4)}=72$, $a_3^{(4)}=96$ and $a_4^{(4)}=24$. The equation for $\p_1(z)$ is
\begin{align}
\hat{A}_4[\p_1](z)&=\Big[-\paz^8-2z^4\paz^4-16z^3\paz^3 +(\om^2-72)z^2\paz^2  \nn\\ 
  &+ (\om^2-2\om E-96)z\paz +\left(-z^8+4\om z^3+E^2-\D^2-24\right)\Big]\p_1=0
\label{redeq4}
\end{align}
The equation for $\g$ reads
\beq
(\g^4+1)^2=0,
\eeq
the exponents of second kind of order two are located on the unit circle. In this case, the two-fold degeneracy of the four different values of $\g$ is not lifted by the exponent of second kind of order one, $\b$, because $\b(\g)=0$ for all $\g$.
The equation determining the exponent of first kind, $\r$, is obtained from the term $b_4$ in the expansion
\beq
\hat{A}_4[\p_1](z)=e^{\frac{\g}{2}z^2+\b z}z^{\r}\sum_{l=0}^\infty b_{8-l}z^{8-l},
\label{expan}
\eeq
and reads
\beq
\r^2+5\r+\frac{\om^2}{16}+\frac{21}{4}=0,
\label{rho4}
\eeq
for all $\g$ with $\g^4=-1$.
The two solutions of \eqref{rho4} are
\beq
\r_\pm=-\frac{5}{2}\pm\frac{\sqrt{16-\om^2}}{4},
\eeq
whose real part is always less than $-3/2$. We conclude again that all formal solutions of \eqref{redeq4} are normalizable and $H_{4p}$ is therefore not self-adjoint. Moreover, the argument of section \ref{sec-k3} about the feasability of possible self-adjoint extensions applies here as well: The Hamiltonian $H_{4p}$ does not describe a physically realizable system.

\section{ The case $\bm{k\ge 5}$ \label{sec-k5}}
We have seen in section \ref{sec-k4} that $\r$ lifts the degeneracy of $\g$ and at the same time determines the normalizability properties of the formal solutions if $\b=0$. We shall show below that $\b=0$ for all $k>4$. Obviously, $\r$ is given for $k\ge 4$ by the condition $b_{2k-4}=0$ in the expansion
\beq
\hat{A}_k[\p_1](z)=e^{\frac{\g}{2}z^2+\b z}z^{\r}\sum_{l=0}^\infty b_{2k-l}z^{2k-l}.
\label{expank}
\eeq
Because the terms depending on $\om$ in \eqref{redeq} contribute at most to the term $b_4$, it is sufficient to consider the reduced operator
\beq
\hat{A}^{red}_k=\paz^{2k}+2z^k\paz^k+a_1^{(k)}z^{k-1}\paz^{k-1}+a_2^{(k)}z^{k-2}\paz^{k-2},
\eeq
which does not depend on $\om,\D,E$. We begin with $b_{2k}=0$ which reads
\beq
(\g^k+1)^2=0,
\eeq
as in the previous cases (in fact, $\g$ may differ form a root of unity only for $k=2$). 

The term $b_{2k-1}$ contains $2k-1$ factors $\g z$
and one factor $\b$ coming  from the operator $\paz^{2k}$, with multiplicity $2k$, likewise the operator $\paz^k$ (from $2z^k\paz^k$) yields $k-1$ factors $\g z$ and one factor $\b$ with multiplicity $k$. The maximal power of $z$ produced by $z^{k-1}\paz^{k-1}$ is $2k-2$, so this part of $\Ar$ does not contribute to $b_{2k-1}$. Apart from these terms in $b_{2k-1}$ multiplying $c_0$ (see \eqref{asym}), there appear terms proportional to $c_1$ which, however, vanish for any $\g$ in the solution set because they are produced by the operator $\paz^{2k}+2z^k\paz^k+z^{2k}$. In a similar fashion, all terms proportional to $c_j$, $j\ge 1$ in $b_{2k-l}$ for $l\le 4$ are redundant because they do not fix the exponents. After the determination of $\g,\b,\r$, the $c_j$ are computed recursively depending on $c_0\neq 0$ which can be arbitrarily chosen.

The terms $\propto c_0$ in $b_{2k-1}$ read
\beq
2k\b(\g^{2k-1}+\g^{k-1})=2k\b\g^{k-1}(\g^k+1),
\eeq
which vanishes due to $\g^k=-1$, so $b_{2k-1}$ does not determine $\b$.

The three operators in $\Ar$ contributing to $b_{2k-2}$ are $\paz^{2k}$, $2z^{k}\paz^{k}$ and $a_1^{(k)}z^{k-1}\paz^{k-1}$. To find the $a_j^{(k)}$, we note the recurrence relation
\beq
a_j^{(n)}=a_j^{(n-1)}+(2n-2j+1)a_{j-1}^{(n-1)}+(n-j+1)^2a^{(n-1)}_{j-2},
\label{recur}
\eeq
with $a_0^{(n)}=1$ and $a_l^{(n)}=0$ for $l<0$. We obtain then
\beq
a_1^{(n)}=\sum_{l=1}^n(2l-1) =n^2,
\label{a1}
\eeq
and
\beq
a_2^{(n)}=2\sum_{l=1}^{n-1}l^3=\frac{(n-1)^2n^2}{2}.
\label{a2}
\eeq
The three relevant operators produce the following terms in $b_{2k-2}$:\\
$\paz^{2k}$:~~ $z^{2k-2}\g^{2k-2}\b^2$ with multiplicity $2k \choose 2$ and $z^{2k-1}\g^{2k-1}\r z^{-1}$ with multiplicity $2k$.\\
$2z^k\paz^k$:~~ $2z^kz^{k-2}\g^{k-2}\b^2$ with multiplicity $k\choose 2$ and
$2z^kz^{k-1}\g^{k-1}\r z^{-1}$ with multiplicity $k$.\\
$a_1^{(k)}z^{k-1}\paz^{k-1}$:~~ $a_1^{(k)}z^{k-1}z^{k-1}\g^{k-1}$ with multiplicity 1.

Apart from the terms proportional to $\b^2$ and $\r$, there are terms coming from the split $\paz^n=\paz\paz^{n-1}$ where $n-1$ operators $\paz$ act on $\exp(\g z^2/2)$ and one $\paz$ on the ensuing factor $\g^{n-1} z^{n-1}$ yielding $z^{n-2}$. Counting all possibilities gives a combinatorial factor of $1+2+3\ldots +(n-1)=n(n-1)/2$.

Collecting all terms gives for the  coefficient of $z^{2k-2}$ in $\left. b_{2k-2}\right|_{c_0}$,
\begin{align}
  &\ph{+}k(2k-1)\g^{2k-1}+k(k-1)\g^{k-1}+a_1^{(k)}\g^{k-1}\nn\\
  &+ \b^2\left({2k \choose 2}\g^{2k-2} +2{k\choose 2}\g^{k-2}\right)
  \label{2k-2}\\
&+(2k\g^{2k-1}+2k\g^{k-1})\r.\nn
\end{align}
The first line in \eqref{2k-2} is $(2k^2-k)\g^{k-1}(\g^k+1)$ because
$a_1^{(k)}=k^2$ and thus vanishes like the last line. Therefore $\r$ is not determined by $b_{2k-2}$. The middle line gives
\beq
k\g^{k-2}((2k-1)\g^k +k-1)\b^2=-k^2\g^{k-2}\b^2.
\eeq
It follows $\b=0$ for all $k\ge 5$.

Setting $\b=0$ from now on, we consider next $\left.b_{2k-3}\right|_{c_0}$. The factor $z^{2k-3}$ has odd exponent and cannot be produced by the even operators in $\Ar$ if $\beta=0$. It therefore vanishes for all $k\ge 5$.

The term $\left.b_{2k-4}\right|_{c_0}$ determines $\r$, but the combinatorial factors are more complicated because the differential operator $\paz^n$ splits now as $\paz\paz\paz^{n-2}$, meaning that two derivatives acting on factors $(\g z)^j$ are interspersed between derivatives acting on $\exp(\g z^2/2)$.

In a first step, we shall derive a generating function for the terms proportional to
$\g^{2k-2}z^{2k-4}$ in $\paz^{2k}\exp(\g z^2/2)$. To this end, we use the representation of $SL_2(\Cc)$ in $\B$ \cite{braak-23}. The generators of $\mathfrak{sl}_2(\Cc)$ are
\beq
K_+=\frac{z^2}{2},\quad K_-=\frac{1}{2}\paz^2,\quad K_0=z\paz+\frac{1}{2}.
\eeq
Then
\beq
\exp\left(\frac{\l}{2}\paz^2\right)\exp\left(\frac{\g}{2}z^2\right)=
\exp(\l K_-)\exp(\g K_+)=\exp(\g' K_+)\exp(\a K_0)\exp(\l'K_-),
\label{expop}
\eeq
with
\beq
\a=-\ln(1-\l\g),\quad \l'=\frac{-\l}{1-\l\g},\quad \g'=\frac{\g}{1-\l\g},
\eeq
which can be obtained via the two-dimensional representation of the group
$SL_2(\Cc)$. We apply now the normal ordered form of the operator in
\eqref{expop} to the constant function and obtain
\beq
e^{\frac{\l}{2}\paz^2}e^{\frac{\g}{2}z^2}=
\frac{1}{\sqrt{1-\l\g}}\exp\left(\frac{\g}{1-\l\g}\frac{z^2}{2}\right)
=G(\l,\g,z).
\label{genf}
\eeq
For a function $f(\g,z)$ one has the expansion
\beq
\exp\left(\frac{\l}{2}\paz^2\right)f(\g,z)=\sum_{n=0}^\infty \l^n\Gamma_n(\g,z),\qquad
\Gamma_n(\g,z)=\frac{1}{n!2^n}\paz^{2n}f(\g,z).
\label{genf2}
\eeq
We expand now $G(\l,\g,z)$ in powers of $\l,\g$ and $z$  (for $|\l|<1$),
\beq
G(\l,\g,z)=e^{\frac{\g}{2}z^2}\sum_{\{n_j\}}\eta_{n_0}\l^{n_0}\g^{n_0}
\frac{1}{\prod_{j=1}n_j!2^{\sum_{j=1}n_j}}
\l^{\sum_{j=1}jn_j}\g^{\sum_{j=1}(j+1)n_j}z^{2\sum_{j=1}n_j},
\label{expanG}
\eeq
where the $\eta_l$ are given as
\beq
\frac{1}{\sqrt{1-\l\g}}=\sum_{l=0}\eta_l(\l\g)^l.
\eeq
We wish to extract the coefficient of the term $\l^k\g^{2k-2}z^{2k-4}$ in \eqref{expanG} which yields the term $C^{(2k)}_0$ proportional to $\g^{2k-2}z^{2k-4}e^{\g z^2/2}$ in $\Gamma_k(\g,z)$ for $f(\g,z)=\exp(\g z^2/2)$. This leads to the conditions
\begin{align}
  n_0+\sum_{j=1}n_j &=k,\nn\\
  n_0+\sum_{j=1}(j+1)n_j &=2k-2,\\
  2\sum_{j=1}n_j &=2k-4.\nn
  \end{align}
They entail $n_j=0$ for $j\ge 4$. The remaining four possibilities for the set $\{n_0,n_1,n_2,n_3\}$ leads to
\begin{align}
e^{-\frac{\g}{2}z^2}\left.G(\l,\g,z)\right|_{\l^k\g^{2k-2}z^{2k-4}} &=
\eta_0\left[\frac{1}{(k-4)!2!0!2^{k-4+2}}+\frac{1}{(k-3)!0!1!2^{k-3+1}}\right]\nn\\
  &+\frac{\eta_1}{(k-3)!1!0!2^{k-3+1}}
+\frac{\eta_2}{(k-2)!0!0!2^{k-2}}.
\label{Cnull1}
\end{align}  
The right hand side of \eqref{Cnull1} gives $C^{(2k)}_0/(2^kk!)$ and we find
\beq
C^{(2k)}_0=\frac{3}{2}k(k-1)+6k(k-1)(k-2)+2k(k-1)(k-2)(k-3).
\label{Cnull2}
\eeq
To compute the combinatorial factor for the operator $\paz^k$ appearing in
$2z^k\paz^k$, we can apply the same formula if $k$ is even (setting $k=2l$).
For odd $k=2l+1$, we obtain instead
\beq
C_0^{(2l+1)}=(2l-2)(2l-1)l+C_0^{(2l)}.
\label{Cnullodd}
\eeq
It turns out that both expressions are the same if written in terms of $k$,
\beq
C_0^{(k)}=\frac{k}{8}(k^3-6k^2+11k-6).
\label{Cnull}
\eeq
The operator $a_1^{(k)}z^{k-1}\paz^{k-1}$ contributes to $z^{2k-4}$ if one $\paz$ acts on $z^{k-2}$, the combinatorial factor is thus $(k-2)(k-1)/2$. The operator $a_2^{(k)}z^{k-2}\paz^{k-2}$ appears for the first time in $b_{2k-4}$, so it has multiplicity 1.
The terms not containing $\r$ in $b_{2k-4}$ are then
\beq
\g^{2k-2}C_0^{(2k)} +\g^{k-2}\left(2C_0^{(k)} +a_1^{(k)}\frac{(k-1)(k-2)}{2}
+a_2^{(k)}\right).
\label{const}
\eeq
The terms proportional to $\r$ and $\r^2$ in $b_{2k-4}$ created by $\paz^{2k}$ are related to two insertions of $\paz$ in the factor $\gamma^{2k-2}z^{2k-2}z^\r$.
If both of them act on $z^\r$, they produce the coefficient $\r(\r-1)$ with multiplicity $2k-2+2 \choose 2$. If only one of the $\paz$ acts on $z^\r$, the other acts on $z^{2k-2}$, yielding the coefficient $\r$. One has to discern two cases, namely whether the first inserted derivative acts on $z^\r$ or the second. In the first case, the combinatorial factor reads
\beq
(2k-1)(2k-2)+(2k-3)(2k-1)(k-1)+\frac{1}{2}\big((k-1)(2k-1)-1\big)
-\frac{1}{2}\sum_{j=2}^{2k-2}j^2,
\eeq
and in the second case
\beq
(2k-1)^2(k-1)-\sum_{j=1}^{2k-2}j^2.
\eeq
Finally, we obtain for the coefficient $C^{(2k)}_{\r^2}$ of $\g^{2k-2}\r^2$ and for the coefficient $C^{(2k)}_\r$ of $\g^{2k-2}\r$,
\beq
C^{(2k)}_{\r^2}=k(2k-1), \qquad C^{(2k)}_\r=4k^3-8k^2+3k.
\eeq
The corresponding factors coming from $\paz^k$ in $2z^k\paz^k$ are computed in the same way to yield the coefficients of $\g^{k-2}\r^2$ and $\g^{k-2}\r$, respectively,
\beq
C^{(k)}_{\r^2}=\frac{k(k-1)}{2}, \qquad C^{(k)}_\r=\frac{1}{2}k(k^2+3)-2k^2.
\eeq
The operator $a_1^{(k)}z^{k-1}\paz^{k-1}$ has a single insertion of $\paz$ in the factor $\g^{k-2}z^{k-2}z^\r$ and produces $\g^{k-2}\r$ with multiplicity $k-1$.
The operator  $a_2^{(k)}z^{k-2}\paz^{k-2}$ does not yield $\r$-dependent terms in $b_{2k-4}$.

Collecting all terms, we find finally for the coefficient of $z^{2k-4}$ in
$\left.b_{2k-4}\right|_{c_0}$,
\begin{align}
&\ph{+} \g^{2k-2}\left[C^{(2k)}_{\r^2}\r^2+C^{(2k)}_\r\r +C_0^{(2k)}\right]\nn\\
&+\g^{k-2}\left[2C^{(k)}_{\r^2}\r^2+\left(2C^{(k)}_\r+(k-1)a_1^{(k)}\right)\r+2C_0^{(k)}
  +\frac{(k-2)(k-1)}{2}a_1^{(k)}
  +a_2^{(k)}\right].
  \label{final1}
\end{align} 
Further simplifications arise when we use now $\g^k=-1$ and \eqref{a1},\eqref{a2}. The final equation for $\r$ has the remarkably simple form
\beq
\r^2+(2k-3)\r+\frac{3}{4}k^2-2k+\frac{5}{4}=0.
\label{final}
\eeq
It has the solutions
\beq
\r_+=-\frac{k}{2}+\frac{1}{2}, \qquad \r_-=-\frac{3k}{2}+\frac{5}{2}.
\label{sol}
\eeq
Both exponents are less than $-1/2$ for $k>2$. Together with the results from sections \ref{sec-k3} and \ref{sec-k4}, it follows that $H_{kp}$ is not self-adjoint in $\B$ for $k\ge 3$. The constructed solutions of the eigenvalue equation are elements of $\mathcal{H}=\mathcal{B}\otimes\Cc^2$, but not of $\mathcal{D}(H_{kp})$, as seen as follows. 
The expectation value of $z^k=\ah^{\dagger k}$,
\beq
|\langle\p_1|z^k|\p_1\rangle| > C_1\int_R^\infty \rd r\ r^k|\p_1(re^{-i\th/2})|^2
=C_2\int_R^\infty \rd r\  r^{k+2\r},
\eeq
with certain positive constants $C_1,C_2$ and $R$ large enough,
is not finite for $\r=\r_+$ and as $\mathcal{D}(H_{kp})\subset\mathcal{D}(\ah^{\dagger k})$, the eigenfunctions must be located in  some (not unique) self-adjoint extension of $H_{kp}$.    
\section{Conclusions \label{sec-con}}
We have shown that the $k$-photon quantum Rabi model is unphysical in the strict sense for $k\ge 3$, because its Hamilton operator $H_{kp}$ is, though apparently hermitian, not self-adjoint. Therefore, according to Stone's theorem \cite{reed-1}, it cannot generate a unitary time development of quantum states. This feature of $H_{kp}$ is not detectable by any numerical evaluation which necessarily operates in a truncated, finite-dimensional Hilbert space. Any such calculation is misleading, even if a calculation of the spectrum seems to converge \cite{ng-99}. The employed method makes use the analytically accessible asymptotics of formal solutions to the eigenvalue equation, although it contains arbitrarily high derivatives and is quite different from the commonly studied case where the ($d$-dimensional) Laplacian is augmented by some potential. Remarkably, the Stokes phenomenon, though certainly present, has no bearing on the conclusions. This is due to the vanishing of the exponent of second kind with order one for $k\ge 4$. If $\b$ does not vanish, it may vary from Stokes sector to Stokes sector. Thus the usual lateral connection problem reappears which is hard to solve by analytic means.
Finally, we  note that the $k$-photon coupling in \eqref{hamk} may acquire physical meaning  in the Dicke models with more than one qubit. Under certain conditions on the parameters of the Dicke models, there exist finite-dimensional invariant sub-spaces with bounded photon content, so-called ``dark-like states''.
The first such state was discovered numerically in \cite{chilingaryan2013} and mathematically identified as exact eigenstate in \cite{peng2014}. Later on, these states have been found in several generalizations of the Dicke model \cite{peng2015} and are highly relevant for applications in quantum information technology because they can be used for the fast generation of $W$-states \cite{peng2021}. For these states to exist, several qubits are required. If the qubit frequencies are different and fine-tuned to the mode frequency (but \emph{not} to the coupling), a condition easily realizable in current exeprimental platforms, there are exact eigenstates of the Dicke Hamiltonian with finite photon number and energy which is independent from the coupling between qubits and the radiation mode \cite{peng2015}. These states bear some resemblance to the quasi-exact states in the asymmetric quantum Rabi model \cite{reyes-bustos2022} and hint at another type of hidden symmetry. Very recently, there has been an attempt to identify this symmetry in the asymmetric generalization of the two-qubit Dicke model \cite{lei2023}. The $k$-photon Dicke model possesses also such exact eigenstates with finite photon number. If restricted to just these states, the Hamiltonian becomes clearly a finite-dimensional diagonal matrix with real entries and is trivially self-adjoint. But the analysis above demonstrates that the generic eigenstates of a system with $k$-photon coupling cannot be meaningfully associated with any self-adjoint extension of its Hamilton operator.

\begin{acknowledgement}
  The author wishes to thank for enlightening discussions with Fumio Hiroshima, Daniel Burgarth and Davide Lonigro.
This work was funded by the Deutsche Forschungsgemeinschaft through grant no. 439943572.
\end{acknowledgement} 


\end{document}